# Enhanced electro-optic modulation in resonant metasurfaces of lithium niobate

Helena Weigand[1*†], Viola V. Vogler-Neuling[1†], Marc Reig Escalé[1], David Pohl[1], Felix Richter[1],

Artemios Karvounis[1], Flavia Timpu[1] and Rachel Grange[1]

[†] These authors contributed equally.

[*] Corresponding Author: hweigand@phys.ethz.ch

[1] ETH Zurich, Department of Physics, Institute for Quantum Electronics, Optical Nanomaterial

Group, Auguste-Piccard-Hof 1, 8093 Zurich, Switzerland

**Abstract** - In display technologies or data processing, planar and subwavelength free-space components suited for flat photonic devices are needed. Metasurfaces, which shape the optical wavefront within hundreds of nanometers, can provide a solution for thin and portable photonic devices, e.g. as CMOS-compatible modules. While conventional electro-optic modulators are inconvenient to operate in free space configurations, its principle can largely be applied to the development of active metasurfaces with the prospect of modulation speeds up to the GHz region. Here, we use this concept to realize fast and continuous modulation of light at low voltage and MHz speed with a lithium niobate metasurface tuned by the linear electro-optic effect. Furthermore, we exploit the resonance in the visible to enhance the modulation of the transmitted light by two orders of magnitude, namely by a factor of 80, compared to the unstructured substrate. This proof-of-concept work is a first important step towards the use of lithium niobate metasurfaces for free space modulation.

**TOC graphic**

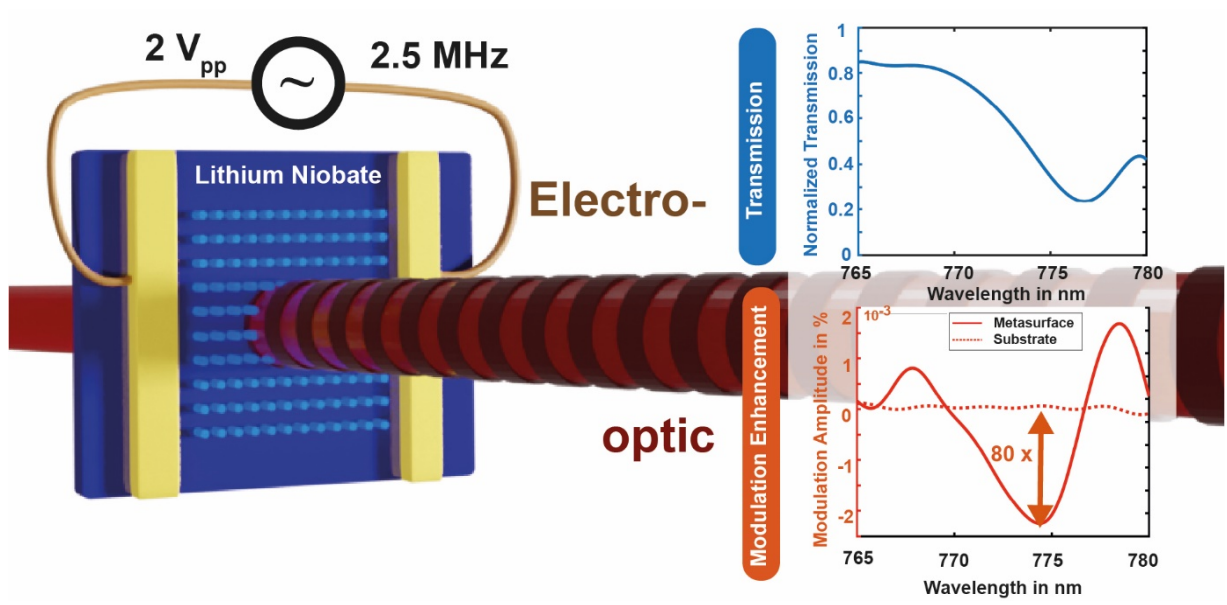

Typical photonic free space applications rely on optical elements such as lenses or polarizers with a thickness of several millimetres. A successful approach for thinner elements are metasurfaces, which influence the electromagnetic wavefront by subwavelength geometries. Although several optical operations have been achieved with metasurfaces[1]–[3] , a current bottleneck in the application is the tunability of their functionality. This issue is tackled by the field of active metasurfaces[4], [5], which is based either on tuning the material properties, unit cell or surrounding medium. Several approaches were demonstrated by exploiting the thermo-optic effect,[6] nonlinear properties,[7]–[9] stretching,[10], [11] MEMS,[12] liquid crystals,[13] plasmonics[14]–[16] or phase change materials.[17], [18] Tuning via electrical actuation reaches the MHz-range and includes approaches such as carrier doping, phase transition, electromechanical or capacitance tuning.[19] Among these, the electro-optic modulation of the refractive index is one of the fastest effects[20] , which is extensively exploited for conventional modulators. Though, light intensity modulation by the linear electro-optic effect has so far not been demonstrated in an all-dielectric metasurface.

LN is an optically transparent, dielectric material widely used in electro-optic modulators,[21]–[23] due to its high refractive index in the visible range ($n_o = 2.26$ at $765$ nm),[24] its high electro-optic coefficient ($r_{33} = 34$ pm/V)[25] and its robustness as inorganic metal-oxide. Recently, lithium niobate metasurfaces were demonstrated and their ability to enhance the second harmonic generation as well as spontaneous parametric-down conversion was shown.[26]–[29]

The electro-optic effect was realized in a metasurface in LN, but it was limited to a phase modulation at 633 nm with 300 $V_{pp}$ and a electrode gap separation of 10 µm at 1 MHz driving frequency.[30] While the Pockels effect is an inherent LN property, here we achieve a wavelength-dependent and 80 times enhanced electro-optic response around the optical resonance of the LN metasurface compared to the substrate. This active electro-optic tuning changes the intensity transmission properties and works up to MHz speeds by applying AC voltages starting from less than 1 V peak-to-peak ($V_{pp}$).

We choose an x-cut LN thin-film (500 nm) on top of a 2 µm silicon dioxide buffer layer and a LN substrate (Fig. 1a). By top-down fabrication, we structure the film into metasurfaces with a period of

500 nm formed of pillar shaped unit cells. The pillars have heights of 200 nm and top radii of 89, 113, 135 and 154 nm for four different metasurfaces. A 300 nm unetched LN layer remains below the pillars. The overall size of a metasurface is 20 x 20 µm². Two gold electrodes with a height of 300 nm, separated by a distance of 25 µm, are used to drive the metasurfaces. The electrodes create an electric field along the $r_{33}$ component of the LN electro-optic tensor (extraordinary z-axis, Fig. 1a).[31]

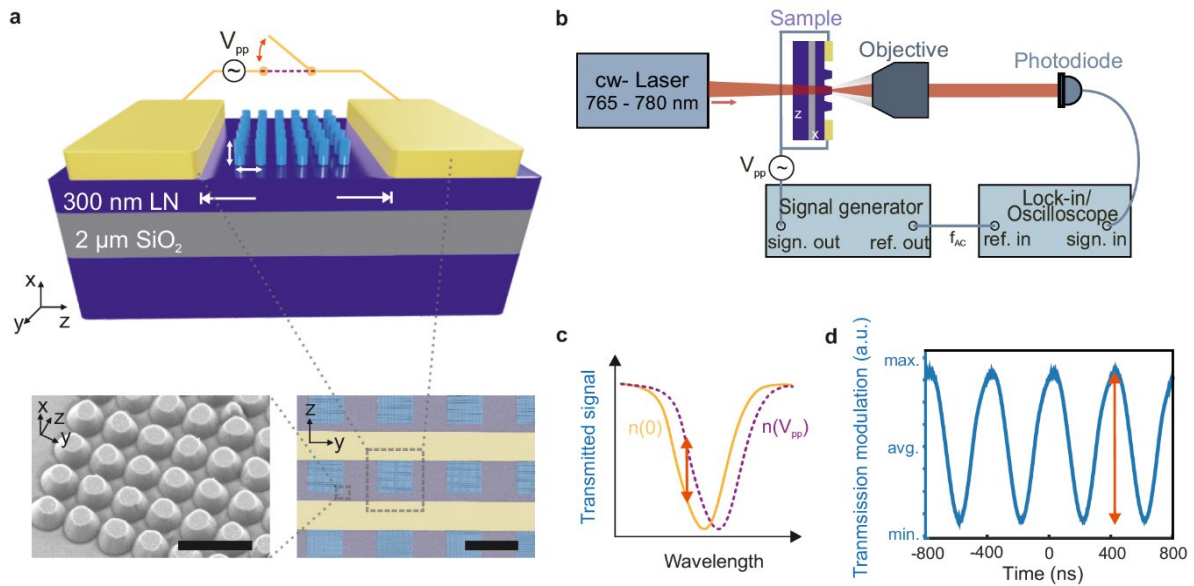

*Figure 1 Design and characterization of an electro-optic metasurface. a Metasurface (blue LN pillars, to differentiate from the remaining LN layer) with pillars of height h and a period p between Au electrodes on top of the sample stack. The lower left inset shows the SEM image of the metasurface pillars structure. Scale bar 1 µm. The lower right inset shows a false-color SEM of several metasurfaces (purple) between the electrodes (yellow). Scale bar 30 µm. b Schematic of the measurement setup. c Schematic transmission curve (yellow line) shifted upon application of a voltage (purple line). The orange arrow marks the change of transmitted intensity i.e. the modulation amplitude. d Measured time trace of the transmitted intensity modulation at 772.4 nm, 22 mW laser power, 2.5 MHz voltage signal and 10 $V_{pp}$.*

Linearly polarized light from a tuneable continuous wave laser, oriented perpendicular to the electrodes, i.e. along the $r_{33}$ direction of lithium niobate, probes the optical behavior of the metasurface with ~0.5 mW optical power. To measure the electro-optic modulation, we apply an AC voltage and monitor the change in transmitted intensity through the metasurface with a lock-in amplifier or an oscilloscope (Fig. 1b, SI Fig. S1). Due to the linear electro-optic effect, the voltage induces a change in the refractive index of LN, which shifts the metasurface resonance, as illustrated schematically in Fig. 1c. The AC voltage modulates the intensity of the transmitted light with the same frequency (Fig. 1d). In this study, we detect this change of transmitted intensity, indicated schematically with an orange arrow with a height named modulation amplitude. We are able to track this electro-optic response using an oscilloscope and report a modulation of the transmitted intensity of 0.01 % for a 180 kHz driving signal with 10 $V_{pp}$. This simple configuration illustrates the broad application range of our device for detection with conventional photodiodes and oscilloscopes compared to e.g. phase modulation.[32]

We determine the resonance position of the metasurface by sweeping the laser between 765 and 780 nm and measuring the transmission intensity with a silicon photodiode with a spectral resolution of 3 pm (Fig. 1b). In Fig. 2a, we show the measured transmission (left axis, solid blue line) of the metasurface with the steepest spectral feature of the four investigated metasurfaces in the given laser range (135 nm pillar radius), normalized to an unpatterned reference area. We observe an optical resonance with a quality factor of 129 at 776.6 nm, where the spectral resonance position can be controlled by the pillar radius (SI Fig. S2). Superimposed on the transmission, we observe Fabry-Pérot interferences with a period of 0.2 nm, originating from the 500 µm thick LN substrate (SI Fig. S3). As the LN substrate is unaffected by the applied electric field, we focus the discussion on the filtered data after the Fabry-Pérot interference removal by Fourier transform filtering (SI Fig. S4).

We measure the modulation amplitude for the wavelength range of the laser while applying an 180 kHz AC voltage of 2 $V_{pp}$. Here, the photodiode signal is tracked by a lock-in amplifier referenced by the signal generator driving the open circuit. In order to determine the enhancement of the modulation as compared to an unstructured film, we define a modulation enhancement factor as the modulation amplitude of the metasurface divided by the modulation amplitude of an unpatterned reference area. The modulation enhancement (Fig. 2a, right axis, solid orange line) shows a strong dependence on the wavelength of the laser, with an about 80 times enhancement in the vicinity of the metasurface resonance. As the unstructured area shows no dependence on the wavelength (SI Fig. S5a Inset) and the electro-optic coefficient of LN is quasi-constant in the investigated wavelength range, this indicates that the modulation amplitude of the structured area is influenced by the transmission of the metasurface itself.

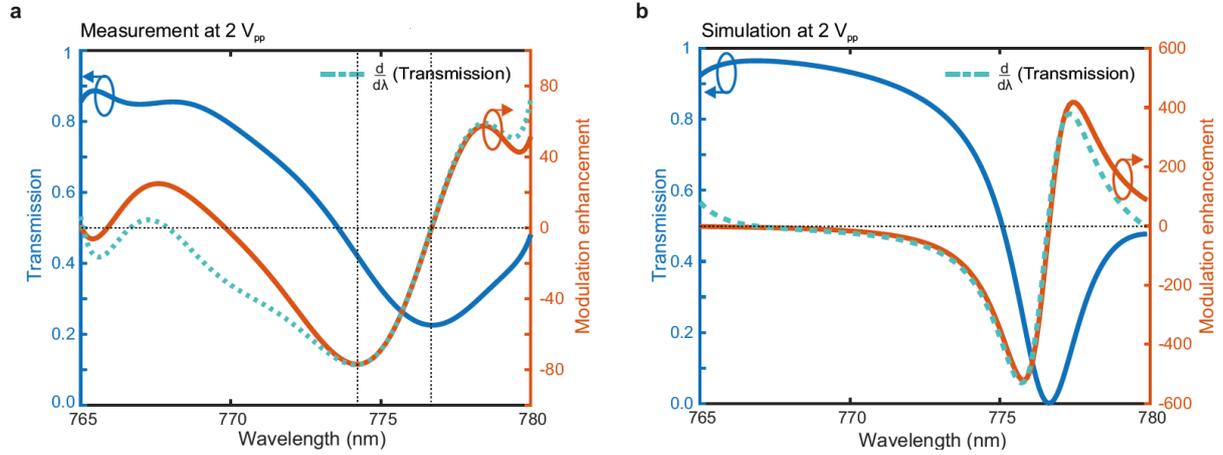

*Figure 2 Wavelength dependence of the electro-optic response. a Measured transmission (blue) of a metasurface with radius 135 nm and period 500 nm, normalized by the transmission of an unstructured area. The orange line shows the modulation enhancement, defined as modulation amplitude of the metasurface divided by the modulation amplitude of an unpatterned area, for an AC voltage of 2 $V_{pp}$ and 180 kHz. The rescaled derivative of the transmitted signal (dashed cyan line) is shown as a guide to the eye. b Calculated transmission (blue) of a metasurface with radius 135 nm and period 500 nm, normalized by the transmission of an unstructured area. The orange line shows the modulation enhancement, defined as modulation amplitude of the metasurface divided by the modulation amplitude of an unpatterned area, for a voltage of 2 $V_{pp}$. The rescaled derivative of the transmitted signal (dashed cyan line) is shown as a guide to the eye.*

As the electro-optic transmission shift is small (SI Fig.S5b inset), we postulate that the modulation amplitude corresponds to the derivative of the transmission. This is validated by the derivative (dashed cyan line) of the measured transmitted intensity of the metasurface (solid blue line) given in Fig. 2a. We conclude that the transmission of the metasurface itself causes a dispersive electro-optic response in the modulation amplitude. This modulation response is therefore unique to every metasurface geometry, as we show for the other investigated metasurfaces (SI Fig. S6). Compared with the response of an unpatterned substrate, which is flat (SI Fig.S5), the modulation amplitude of the metasurface is enhanced by two orders of magnitude ($-2.2 \cdot 10^{-3}$ % on the metasurface compared to $-0.026 \cdot 10^{-3}$ % on the substrate) at 774.2 nm (dashed line in Fig. 2a as guide to the eye). These modulation amplitudes correspond to an about 80 times modulation enhancement at the metasurface compared to the substrate. It is worth noting that the negative sign of the enhancement factor or modulation amplitude does not imply less modulation (this would be true for factors between 0 and 1 such as at the resonance itself).

Subsequently, we investigate the origin of the electro-optic modulation enhancement with a finite element method (FEM) model. The transmission of the metasurface (Fig. 2b, left axis, solid blue line), normalized to the reference layer (SI Fig. S5b inset), was calculated using a model described in SI S5 Fig. S7 and matches the measurements (Fig. 2a). The quality factor of 272 of the simulated transmission resonance is higher than the measured one (129), which we attribute to fabrication imperfections. In the next step, we calculate the electric field distribution created by the gold electrodes using an extended model described in SI Fig S10. Owing to the electrodes geometry, the electric field in the metasurface takes higher values in the edge unit cell and decreases uniformly towards the center of the metasurface (SI Fig S10, insets). Using the simulated local electric field distribution in the central unit cell, we determine the local refractive index modification caused by the electro-optic effect. With the new value of the refractive index, we determine the modulation at 2 $V_{pp}$ (Fig. 2b, right axis, orange curve) using the method described in SI Section S8. Our simulation confirms the experimental result that the modulation of the metasurface has the highest enhancement (and absolute amplitude, Fig. S5) at the steepest slope of the transmission spectrum. The deviation in magnitude between experimental and numerical enhancement factor is attributed to less pronounced edges between the pillars due to fabrication imperfections.

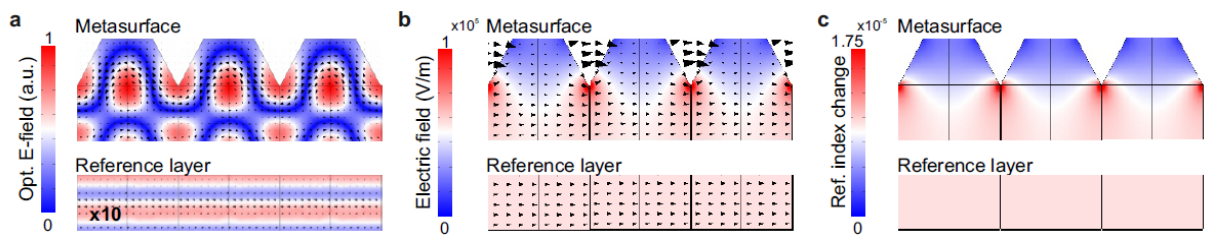

*Figure 3 . Optical field, electric field and refractive index change distribution over metasurface a* Optical field strength at the resonance wavelength in one unit cell of the metasurface (right) and reference layer (left) along a plane normal to the incident electric field. The arrows show the direction of the magnetic field. *b* Transversal cut with the distribution of the electric field norm created by the electrodes for the metasurface (up) and the reference layer (bottom). The arrowheads show the direction of the electric field. The cut is transversal to the LN surface, with the extraordinary axis along the vertical. *c* The change of refractive index induced by the electro-optic effect at 1 V in the metasurface unit cell (up) and the reference layer (bottom). In (a-c), three unit cells (width of 500 nm) of the metasurface/reference layer is shown. The change of refractive index is largest at the corners of the pillars as it is the position where the optical electric field (*a*) and the electric field (*b*) are largest.

The optical field strength map in Fig. 3a shows that the resonance is dominated by the electric dipole mode (SI Fig. S9). Further, we calculate the electric field from the electrodes with an FEM model (SI Fig. S10). We observe that the metasurface nanopatterning imprints a periodicity on the electric field (Fig. 3b, top), whereas the electric field in the reference layer is uniform (Fig. 3b, bottom). As expected for a dielectric material, the electric field strength is larger in the surrounding air (Figure S11). As air does not exhibit an electro-optic response, the electric field strength in color is neglected in Figure 3b due to better visibility. Comparing Fig. 3a and b, a clear overlap of the electric hotspots with the optical field at the corners of the metasurface is visible. This is relevant as the electric field hotspots are also the locations of biggest change in refractive index (Fig. 3c), which will eventually change the transmission behavior of the metasurface. On average, the refractive index change has similar values for the metasurface and the reference layer. However, around the resonance this change leads to a much stronger effect on the transmission and the electro-optic modulation. Having investigated the microscopic origin of the modulation, the next section will be dedicated to the evaluation of the performance of the device and the characterization of the linear electro-optic effect.

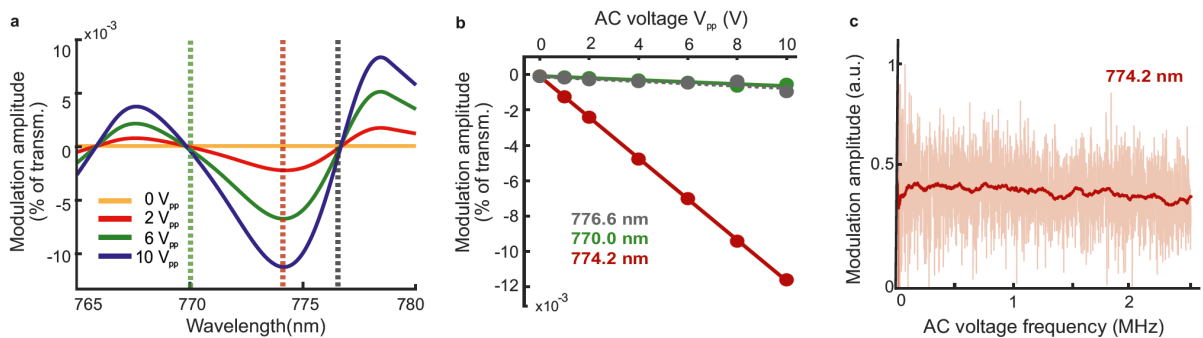

*Figure 4 Performance of the electro-optic metasurface. **a** Dependence of the modulation amplitude on peak-to-peak AC voltage between 0 $V_{pp}$ (no modulation) and 10 $V_{pp}$ (strong modulation) with 180 kHz driving frequency. The dashed coloured lines are the same as in Fig. 2a and indicate the position of the cross-sections in b. **b** Modulation amplitude upon application of an increasing AC voltage at 770 nm (green line), 774.2 nm (red line) and 776.6 nm (grey, dashed line). The peak-to-peak voltage is indicated on the horizontal axis. The dots are measured data at 180kHz driving frequency, while the solid lines are linear fits. **c** Measured frequency dependence of the modulation amplitude between 1 kHz and 2.5 MHz at 774.2nm. The solid red line is the smoothed data.*

Conveniently, the dispersive electro-optic modulation response offers two approaches for changing the electro-optic modulation strength: the operating wavelength and the driving voltage. The modulation amplitude changes the most at the extrema of the derivative of the transmission, as shown in Fig. 4a for different $V_{pp}$ values. Consequently, the electro-optic modulation barely changes at the extreme values of the transmission itself (green and grey dashed line), even upon application of higher voltages. To investigate theinfluence of the electric field on the modulation amplitude, we fix the laser at chosen wavelengths (Fig. 4b): off-resonance (770 nm), at the steepest drop of the transmitted intensity (774.2 nm), and at the resonance (776.6 nm). Upon increasing $V_{pp}$, we observe that the modulation amplitude scales linearly with the applied voltage and reaches 0.01 % at 10 $V_{pp}$. This strongly validates the dominance of the linear electro-optic effect for the modulation of the metasurfaces and is further confirmed by the independence of the modulation amplitude on the laser power, that excludes a thermo-optic effect (SI Fig. S12). The electro-optic modulation can be detected already for voltages below 1 $V_{pp}$, indicating the practical impact of this device for low-power consumption applications. The different slopes in Fig. 4b reveal that the modulation at 774.2 nm scales one order of magnitude higher to the applied voltage than the off-resonant/on-resonance case, which further confirms the dispersive electro-optic behaviour.

The bandwidth of the modulation, which defines the driving frequencies of the device, is in the MHz region. We can operate in a frequency range from a few Hz (SI Fig. S12) up to the MHz-region, with a flat response to the driving frequency from 1 kHz up to 2.5 MHz (Fig. 4c), limited only by the electrical scheme. Based on LN properties, our device should be operational at modulation frequencies up to the GHz regime.[22], [33] However, the modulation response in the MHz-frequency range already supports the dominant linear electro-optic effect since other modulating effects (e.g. thermal) exhibit lower characteristic timescales.[34]

In conclusion, we demonstrate an all-dielectric LN metasurface in the visible with a resonance that can be actively and continuously tuned based on the linear electro-optic effect. This metasurface shows a versatile way of realizing small, low-loss and tuneable optical devices operating in the visible to near infrared range and offering fast modulation. We show that the electro-optic modulation amplitude is wavelength-dependent and enhanced through the optical resonance of the metasurface by a factor of 80 in magnitude compared to an unstructured film, which corresponds to an two orders of magnitude enhancement. The agreement with the simulations supports the experimental findings and their interpretations. We achieve a frequency-independent modulation amplitude between 10 Hz and 2.5 MHz for voltages below 1 V, which allows driving our device by commercial micro-controllers. The modulation amplitude of 0.01 % in our proof-of-concept work is rather small for practical applications. There are several strategies on how to boost the modulation amplitude in the future. First of all, a sharper resonance with a higher Q-factor would increase the modulation amplitude. Promising resonance types could be for example bound states in the continuum or Fano resonances.[35–37] Enhancing the overlap of the electric with the optical field would improve the modulation, which could also further be optimized by an advanced fabrication to obtain sharper structures. Electrodes on top and on the bottom of the metasurface may be preferable, but such a configuration is a challenge with the current available LN substrate platforms. Another interesting approach is to combine LN with other all-dielectric materials e.g. silicon for hybrid electro-optic platforms as theoretically predicted.[38]

Future work includes an extension of the setup to investigations up to the GHz-regime and an optimized design for better electric and optic field overlap or higher Q-factor resonances. This could lead to a stronger enhancement of the electro-optic modulation amplitude. However, our proof-of-concept work already gives the experimental demonstration on how the electro-optic modulation amplitude is enhanced at a metasurface resonance and is to date the fastest and strongest electro-optically modulated metasurface. With further improvements, tuneable lenses and color displays are within reach. The fast tuning properties of our LN device could lead to advances in beam steering and ultimately push free-space data processing to the next level.

Methods

**Metasurface fabrication**

The metasurfaces presented in this work were fabricated at the cleanroom facilities of the Binnig and Rohrer Nanotechnology Center (BRNC). The sample was fabricated on a commercially available x-cut lithium niobate-on-insulator thin film chip with a top-down approach using reactive ion etching. The gold electrodes were deposited using a lift-off process after electron-beam evaporation. A detailed fabrication report can be found in the Supporting Information.

**Simulations**

All the presented simulations were performed using FEM models in COMSOL Multiphysics 5.5, which are described in detail in the Supporting Information. We modeled the origin of the rapid oscillations in Fig. 1e by comparing two 2D FEM models, one including the full 500 µm LN wafer and one including a semi-infinite LN wafer (SI Section S3). We calculate the transmission of a metasurface and of the reference using a 3D model shown in Section S6, which takes into account the particular geometry of our sample. This method is based on previous works, described in detail elsewhere [39]. To account for the electro-optic effect, we combine the method shown in Section S6 with a previous step using the Electrostatic module of COMSOL Multiphysics (SI Section S8). In this step, we calculate the electric field created by two gold electrodes in a full slice along the extraordinary axis. Using this result, we update the value of the refractive index with the electro-optic contribution and use the new refractive index value to run the model in Section S6.

**Acknowledgments:**

We acknowledge support for nanofabrication from FIRST of ETH Zurich. We thank the Cleanroom Operations Team of the Binnig and Rohrer Nanotechnology Center (BRNC) for their help and support. This work was supported by the Swiss National Science Foundation Grant 179099, the European Union's Horizon 2020 research and innovation program from the European Research Council under the Grant Agreement No. 714837 (Chi2-nano-oxides), FP-RESOMUS, the Swiss National Science Foundation through the NCCR MUST. This work was also supported by Swiss National Science

Foundation grant no. 150609. F.T. and H.W. acknowledge financial support from the Physics Department at ETH Zurich.

**Supporting Information**

Supporting Information Available:

Detailed description of fabrication, simulation, and experimental setup. Investigation of other metasurfaces with different resonances and study on laser polarization dependence and laser power dependence. Multipole expansion of the resonance, investigation of Fabry-Perot-resonances and simulations on the electric field distribution are presented. Modulation at low frequencies is shown. This material is available free of charge via the Internet at http://pubs.acs.org

**Contributions:**

R.G., F.T., H.W. and V.V.-N. designed the experiment. V.V.-N., H.W. and F.R. built the setup. H.W., V.V.-N. and F.T. conducted the experiments. M.R.E., H.W. and D.P. fabricated the device. H.W analysed the data with contributions from F.T. and V.V.-N. F.T. did the simulations. H.W., F.T and V.V.-N. wrote the manuscript with contributions from M.R.E.,D.P., F.R., A.K. and R.G.

# Supporting Information

# Enhanced electro-optic modulation in resonant metasurfaces of lithium niobate

Helena Weigand[1*†], Viola V. Vogler-Neuling[1†], Marc Reig Escalé[1], David Pohl[1], Felix Richter[1,] Artemios Karvounis[1], Flavia Timpu[1] and Rachel Grange[1]

† These authors contributed equally.

*Corresponding Author

[1]Optical Nanomaterial Group, Department of Physics, ETH Zurich, Auguste-Piccard-Hof 1, 8093 Zurich, Switzerland

†These authors contributed equally.

*Corresponding author: **hweigand@phys.ethz.ch**

## Contents

# S1 Electro-Optic Setup

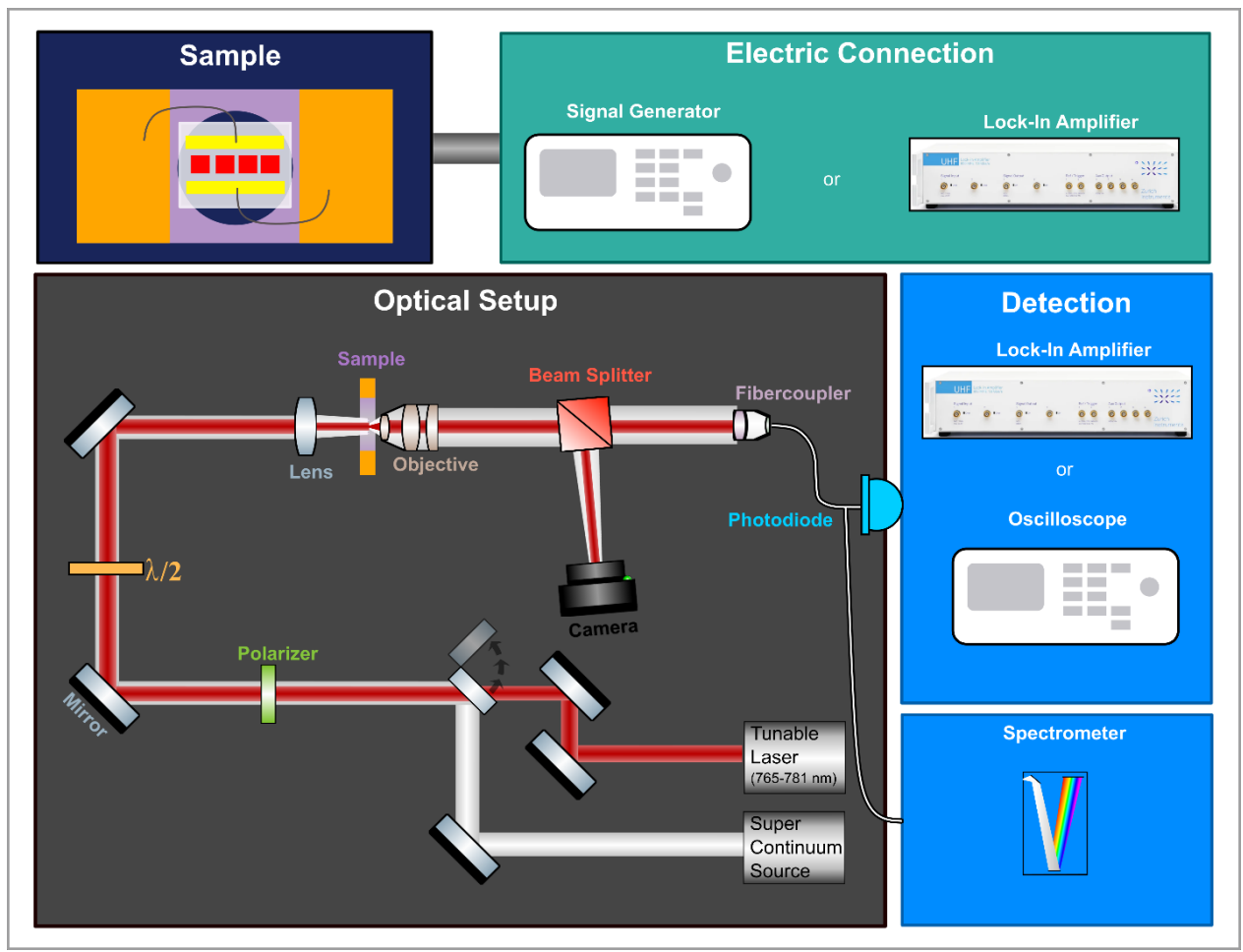

*Figure S1: Electro-Optic Setup*

In **Figure S1** we show a detailed schematic of the electro-optic setup. Two sources of excitation can be used: either a super continuum source for broad wavelength illumination or a tuneable continuous wave laser operating between 765 nm and 781 nm. With the polarizer, we control that both sources have the same incident polarization angle. The incident polarization on the sample is controlled with a half-wave plate. A lens with a focal length of 50 mm is slightly focussing the beam onto the sample. The size of the excitation spot is 50 µm in diameter. We collect the signal from the sample with a 50x collection objective and a fibre collimator plus fibre on a photodiode or for the spectral measurements directly on the imaging spectrometer. With a backwards illumination through the fibre that acts as a pinhole, we make sure that we are indeed collecting the signal from the metasurface. The signal on the photodiode is processed either on a lock-in amplifier or on an oscilloscope. The sample possesses gold electrodes as described in the Methods section of the paper. The gold pads are connected to the electrical pads on the sample holder by silver paste. Two crocodile clamps are clipped on these pads and are connected to a BNC cable. The electrical driving is either done with a frequency generator ($V_{pp}$) or with a lock-in amplifier.

## S2 Spectral Measurements of Characterized Metasurfaces

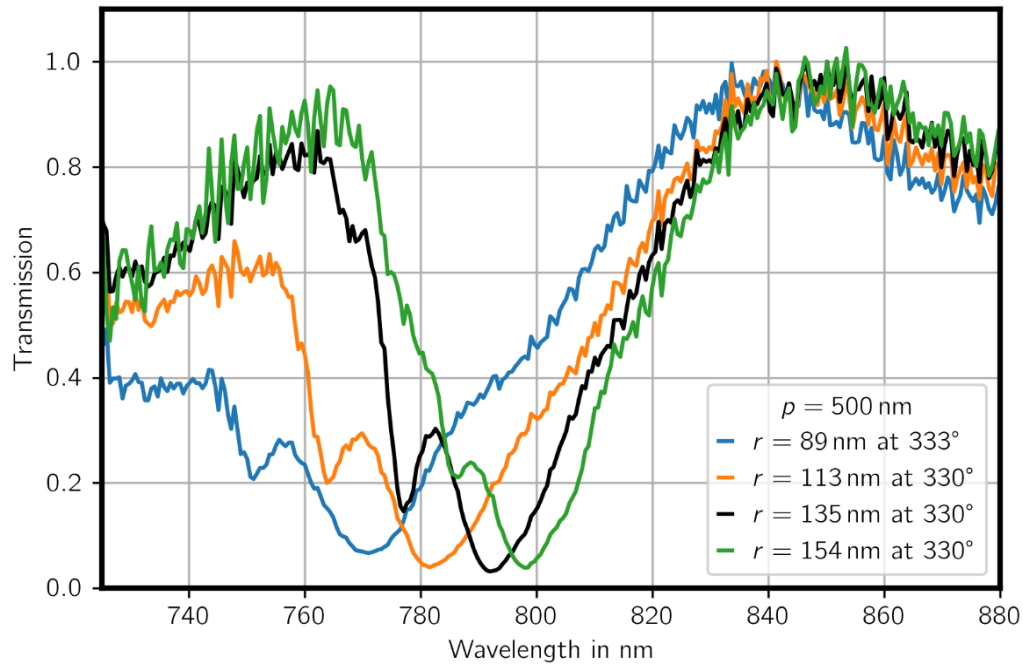

*Figure S2: Transmission measurements of characterized metasurfaces. The spectra of four different metasurfaces all with a period of 500 nm and four different radii (89 nm, 113 nm, 135 nm and 154 nm) were measured with a super continuum source and an imaging spectrometer. The resonance positions redshifts with larger radii.*

We preliminary characterized four metasurfaces with a period of 500 nm and different radii (89 nm, 113 nm, 135 nm and 154 nm) (**Figure S2**). The metasurfaces were excited with a super continuum source and their response was collected with an imaging spectrometer (**Figure S1**). The redshift of the resonance for larger pillar radius is a clear indication that we are in the subdiffractive metasurface regime. The polarization value of 330° correspond to the direction perpendicular to the electrodes and also to the value that gives the steepest slope in the resonance, which is desired for the electro-optic measurements. The metasurface with radius of 135 nm (Figure S2, solid black line) has the resonance well within the CW laser range.

## S3 Origin of the Measured Fast Oscillation in the Transmission

The measured transmission of the LN metasurface shows a fast oscillation superimposed on the transmission resonance. We verify the origin of this oscillation in a 2D FEM model, in which the LN wafer has a finite thickness of 525 µm. The unit cell is embedded in air and has periodic boundary conditions. The 2D model is preferred here due to computation limitations.

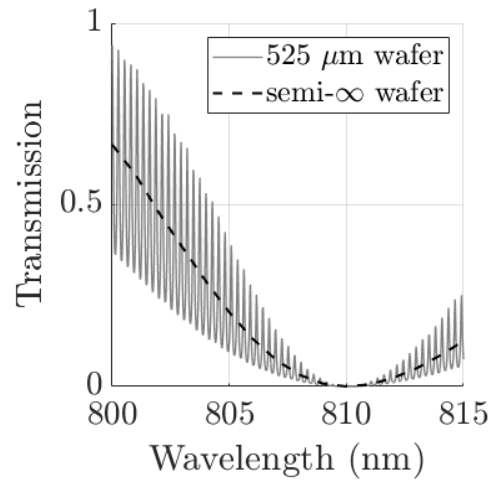

*Figure S3: Calculated optical transmission of a LN metasurface* placed on a LN wafer with a finite thickness (solid grey line). For comparison, the optical transmission of a LN metasurface placed on a semi-infinite LN wafer is also calculated (dashed black line).

The calculated optical transmission of the metasurface with finite LN wafer in **Figure S3** (solid grey line) and a semi-infinite LN wafer **Figure S3**, dashed black line) are shown. The simulation clearly shows that the oscillation is a Fabry-Pérot interference occurring in the LN wafer. The period of this oscillation is 0.27 nm, in very good agreement with the measured value of 0.2 nm.

An experimental evaluation of the transmission data in form of a Fourier transform confirms the dominant frequencies of the Fabry-Pérot interference and allows a filtering of the data to remove these oscillations, as shown in **Figure S4**. The data were thus analyzed by Fourier filtering and subsequent fitting as a $9^{th}$ degree polynomial for the sake of having a meaningful derivative of the transmission (used for comparison to the modulation).

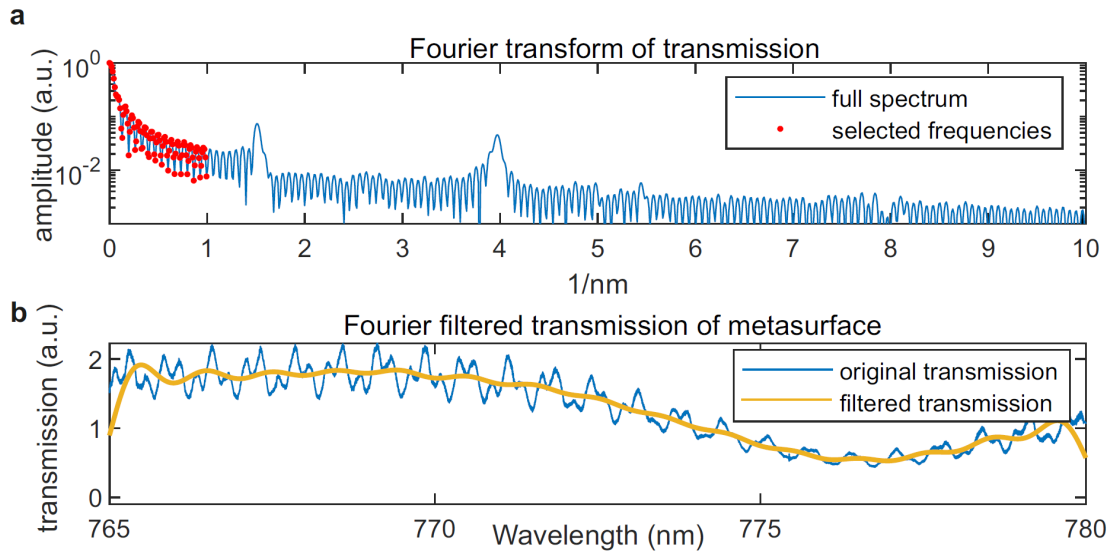

*Figure S4: a) Fourier analysis of metasurface transmission* for the metasurface with a period of 500 nm and a radius of 135 nm. *b) The reconstruction of the signal (yellow line) is based on the frequencies marked with red points (Fig. 4a), thus excluding the dominant oscillations visible in the raw signal (blue line).*

## S4 Total Modulation Amplitude in Experiment and Simulation

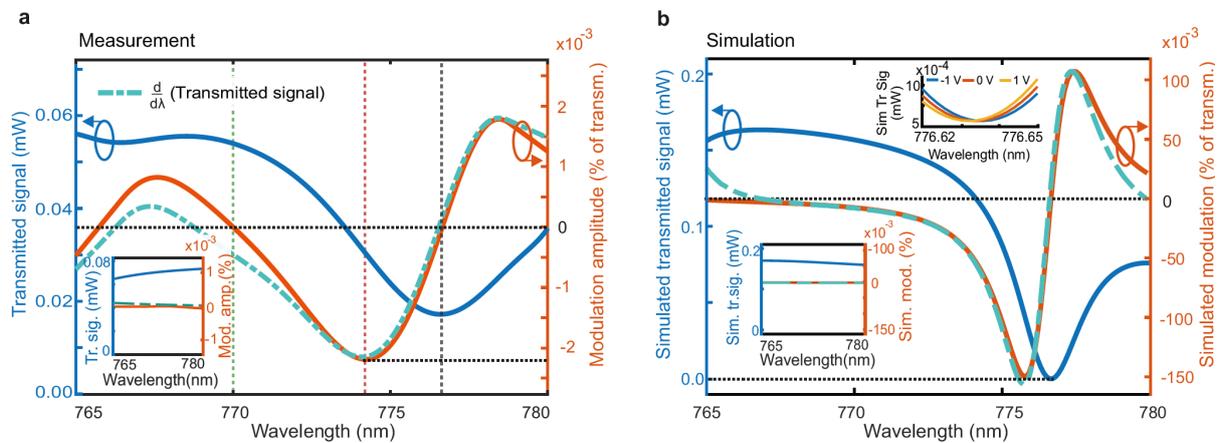

*Figure S5  a Measured transmitted signal (blue) of a metasurface with a radius 135 nm and period 500 nm and the modulation amplitude for an AC voltage of 2 $V_{pp}$ and 180 kHz (orange line) as percentage of the maximum transmission signal (extinction ratio). The rescaled derivative of the transmitted signal (dashed cyan line) is shown as a guide to the eye. Inset: the same measurement on a reference layer (300 nm LN thin film). b Calculated transmitted signal of the LN metasurface (solid blue line) using an FEM model and its derivative (dash-dotted cyan line). Calculated modulation amplitude of the transmission for an AC voltage of 2 $V_{pp}$ (solid orange line) as percentage of maximal transmission. The derivative is rescaled to the maximum value of the transmission signal. The bottom inset shows the same calculations for the reference layer. The top inset shows the simulated transmitted signal by the metasurface around the resonance for -1V (blue), 0 V (orange) and 1V (yellow) bias*

## S5 Electro-optic Modulation for other Metasurfaces

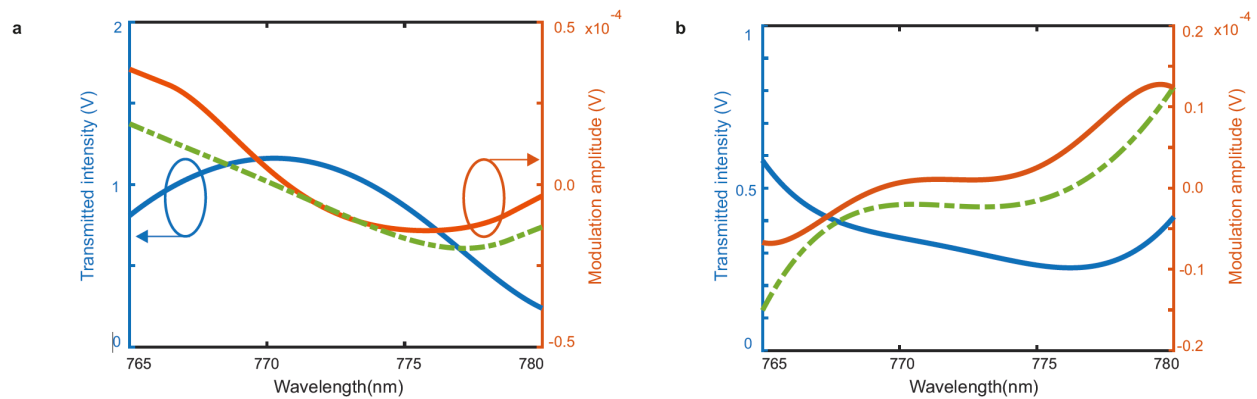

*Figure S6: **Measured electro-optic modulation** of metasurfaces with a 113 nm radius and b 89 nm radius. The blue line represents the transmission, the orange line the modulation amplitude and the green dashed line the derivative of the transmission, scaled as a guide to the eye.*

To show the dependence of the modulation on the geometry of the unit cell, we investigated metasurfaces with different pillar radii as well. Although the transmission resonance itself is smaller, one can still clearly see the dependence of the modulation amplitude on the shape of the transmission. This becomes especially clear when comparing the derivative of the transmission (green dashed line) with the modulation amplitude. Note that the measurements were taken at the respective best polarization for the metasurface, optimized for the steepest drop in transmitted intensity (333° for 89 nm radius and 283° for 113 nm radius).

## S6 Calculation of the LN Metasurface Optical Transmission

We calculate the optical transmission of a LN metasurface using an FEM model (COMSOL Multiphysics 5.5). To show that the transmission contains a resonance which is characteristic to a metasurface, we compare it to the optical transmission of a reference LNOI wafer with a thickness of 300 nm.

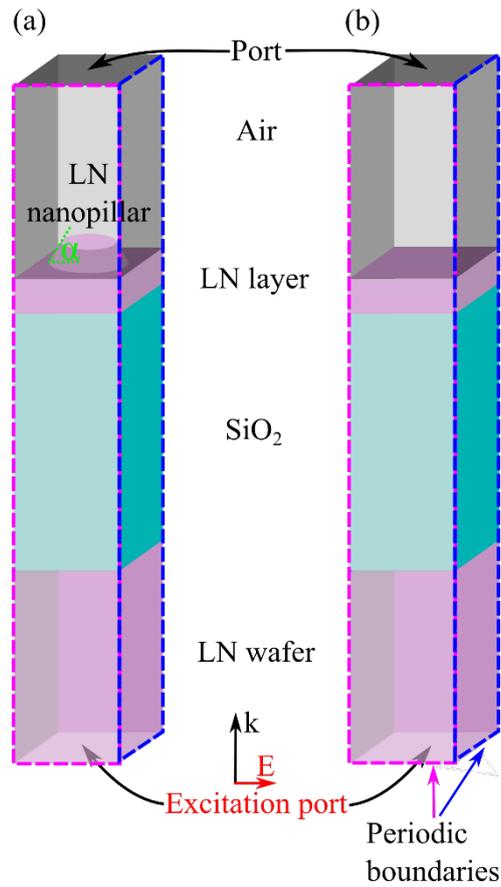

*Figure S7: Schematic of the FEM model used to calculate the optical transmission of a LN metasurface (a) and of a reference LN layer (b). The reference LN layer has a thickness of 300 nm. The $SiO_2$ layer is 2 µm thick. Then angle α is the etch angle characteristic for the ICP fabrication process. The incident light is polarized along the extraordinary axis of LN.*

The FEM models is shown schematically in **Figure S7** and consists of unit cells combined with periodic boundary surfaces. In the metasurface unit cell, the tilt angle α of the nanopillar wall (**Figure S7a**) appears due to the ICP etching process. The bottom port is used to model the incident light, polarized along the extraordinary axis of LN. The top port is used to absorb the transmitted light. No diffraction orders are found in the chosen wavelength range of our measurements.

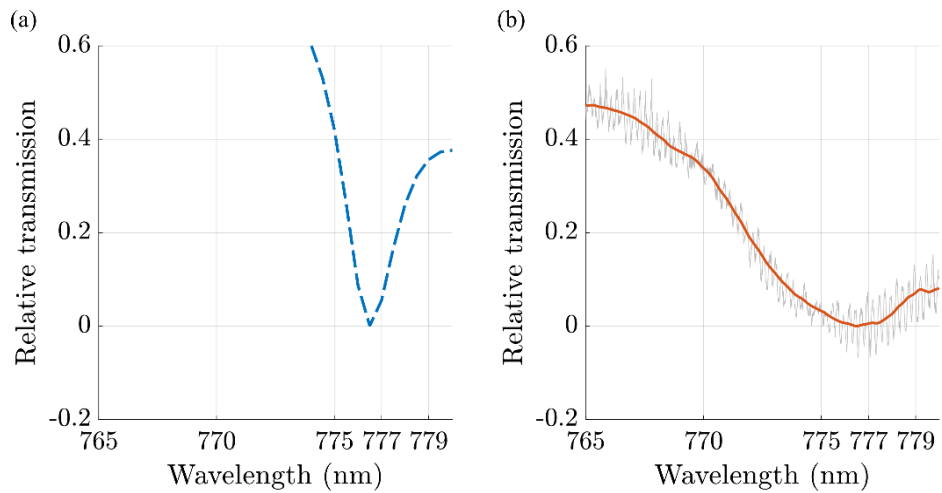

*Figure S8: Optical transmission of an LN metasurface relative to the reference layer. The criterion for the fit is the position of the transmission resonance. The measured relative transmission (b) shows the metasurface resonance with superimposed Fabry Perot interferences, whose origins is discussed in section S3.*

In **Figure 8a** we show the best fit for the calculated optical transmission of the LN metasurface. For this calculation we chose an etch angle of 62°, a nanopillar height of 230 nm and a top radius of the nanopillar of 126 nm. All these values are within the confidence interval of values deduced from SEM and AFM images of the fabricated metasurfaces. The refractive index of LN was taken from the literature.[S1]

The calculated transmission resonance has a higher quality factor in comparison to the measurement. This difference appears due to fabrication imperfections and mismatches between the model and the measurement, as already discussed in details in previous works.[S2,S3]

## S7 Multipole Decomposition

We calculate the multipole expansion (**Figure S9**) of the metasurface from the optical field in the metasurface, using a method already described in detail in previous works.[S4][1] Even though that there are several multipole present the resonance is mainly originating from the electric dipole.

---

[1] Comsol: Multipole Analysis of Electromagnetic Scattering (Application ID: 31901), https://www.comsol.de/model/multipole-analysis-of-electromagnetic-scattering-31901, accessed on 16th of June 2021

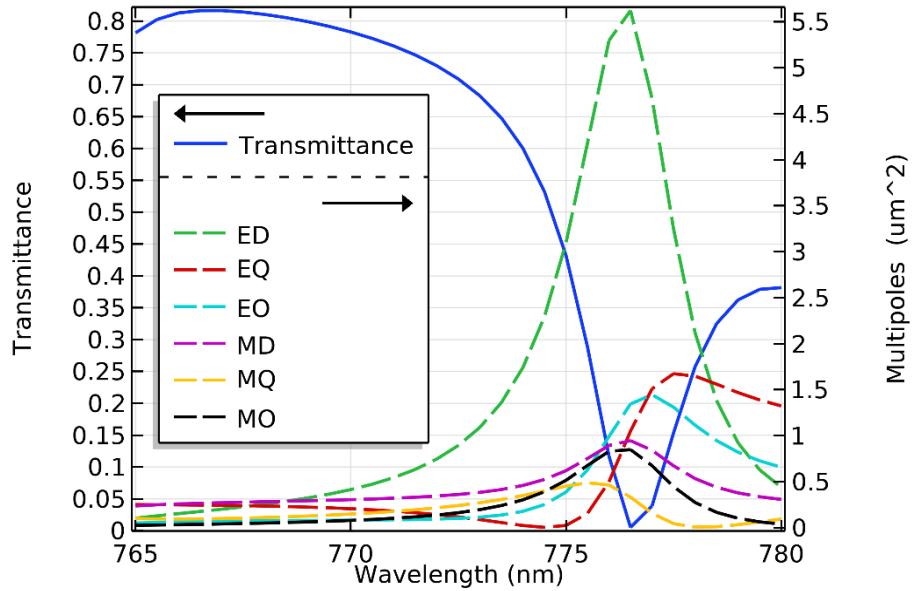

*Figure S9: Multipole expansion of optical field inside the metasurface.* The solid blue line shows the total transmission of the metasurface and the dashed lines show the contributions from: electric dipole(ED) – green, electric quadrupole(EQ) – red, electric octopole(EO) – cyan, magnetic dipole(MD) – magenta, magnetic quadrupole(MQ) – orange, magnetic octopole(MO) – black.

## S8 Simulation of the Electro-Optic Response of a LN Metasurface

We use an FEM model combining electrostatic and frequency domain calculation to estimate the change of the optical transmission when an ac voltage is applied to the electrodes. For this model, we make several approximations, as described in the following.

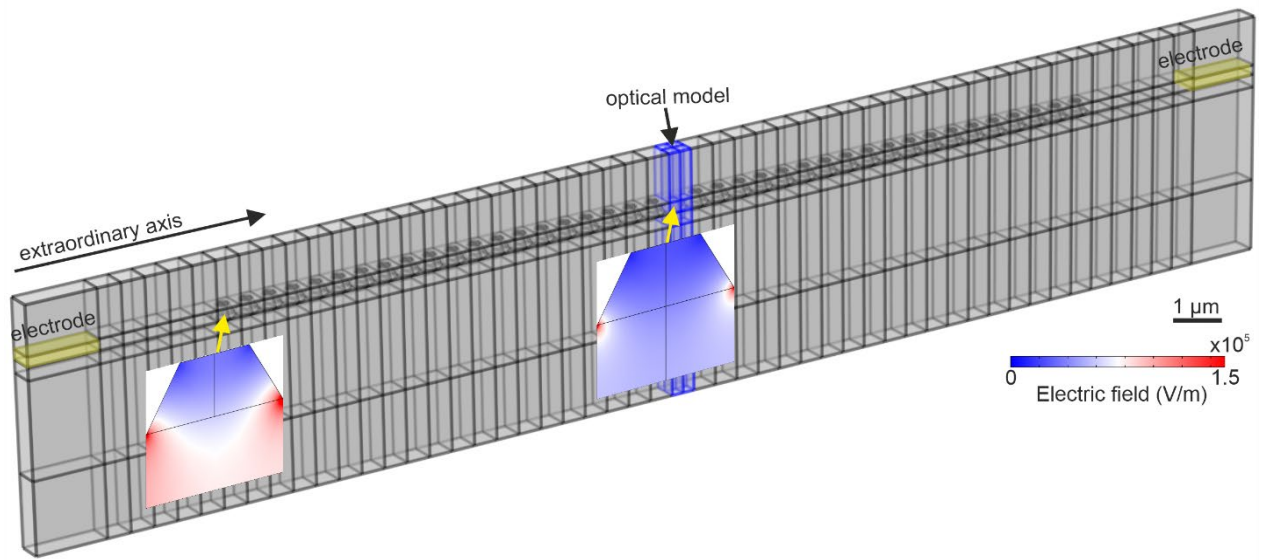

*Figure S10: FEM model used to determine the electro-optic modulation* of the optical transmission. The electrostatic field created by the yellow coloured electrodes is calculated for the entire model. The optical transmission is calculated using the

*blue coloured part of the model. The two insets show the electric field in the first and the middle unit cell of the metasurface.*

The first approximation concerns the size of the model. By using the symmetry of the electrodes, we reduce the model to a slice along the extraordinary axis containing an array of 40 unit cells and two electrodes (**Figure S10**). Perpendicular to the extraordinary axis, we employ periodic boundary conditions.

For the second approximation, we define the change of the refractive index due to the electro-optic Pockels effect as:

$$\Delta n(\lambda, x, y, z) = -\frac{1}{2} r_{33} n_e^3(\lambda) E_e(x, y, z) \qquad (1)$$

where λ is the wavelength, x, y, z are coordinates, $n_e$ is the extraordinary refractive index in the absence of the voltage bias and $E_e$ is the electric field created between electrodes along the extraordinary axis. This approximation is suitable because the electric field from the electrodes is parallel to the extraordinary axis of LN, which also has the highest electro-optic tensor components.

The optical transmission of the metasurface when the voltage $V_0$ is applied is calculated as already shown in Fig. S5, but with a different refractive index of LN, defined as:

$$n_e^{V_0}(\lambda, x, y, z) = n_e(\lambda) + \Delta n(\lambda, x, y, z) \qquad (2)$$

$$n_o^{V_0}(\lambda) = n_o(\lambda) \qquad (3)$$

For the calculation of $\Delta n(\lambda, x, y, z)$, we use the electric field calculated in the 20[th] unit cell, in the middle of the metasurface (**Figure S10**). The electric field in the middle of the metasurface has the lowest value (**Figure S10**, inset), therefore this approximation does not overestimate the electro-optic effect. We use the same approach for the reference layer transmission.

## S9 Electric Field in Air Surrounding the Metasurface

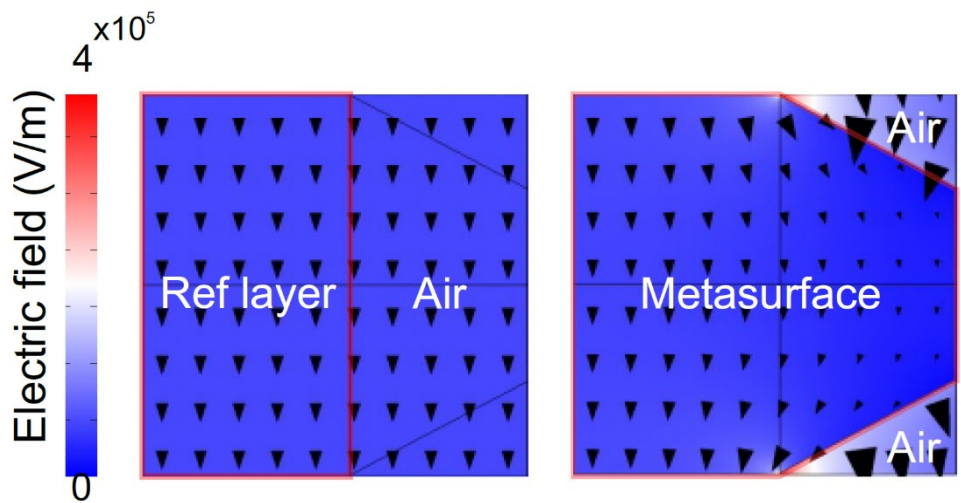

*Figure S11 Electric field strength over the sample geometry* The electric field strength in the air is higher than in the dielectric structure, which is expected from theory. The color code in the main manuscript of the air layer was neglected because air does not show an electro-optic response and is therefore not important for the discussion. Furthermore, the electric field differences within the dielectric are not visible if the color of the air area is present.

## S10 Power Dependent Measurements on Metasurface

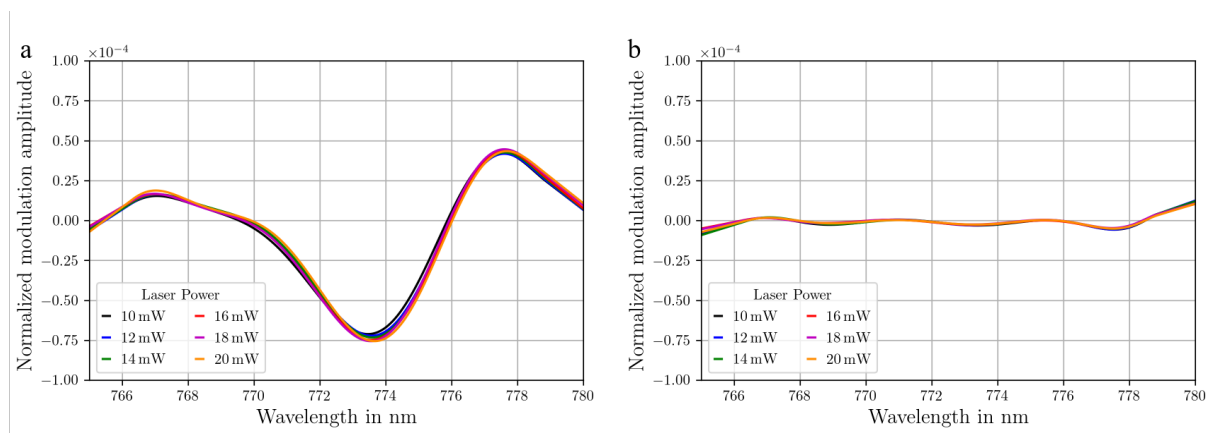

*Figure S12: Power Dependent Measurements* for six laser powers 10 mW, 12 mW, 14 mW, 16 mW, 18 mW and 20 mW for (a) a metasurface with period 500 nm and radius 135 nm and (b) a reference layer with a thickness of 300 nm.

To exclude that our modulation is based on the photorefractive effect, we measured the modulation amplitude of a metasurface of a period of 500 nm and a pillar radius of 135 nm and of a LN substrate of 300 µm thickness for six different laser powers (**Figure S11**). The modulation amplitude was normalized with the respective maximum intensity value. The spectra for different laser powers do not differ for different laser powers. Therefore, our modulation amplitude is not due to a photorefractive effect.

## S11 Frequency Response at 10 Hz

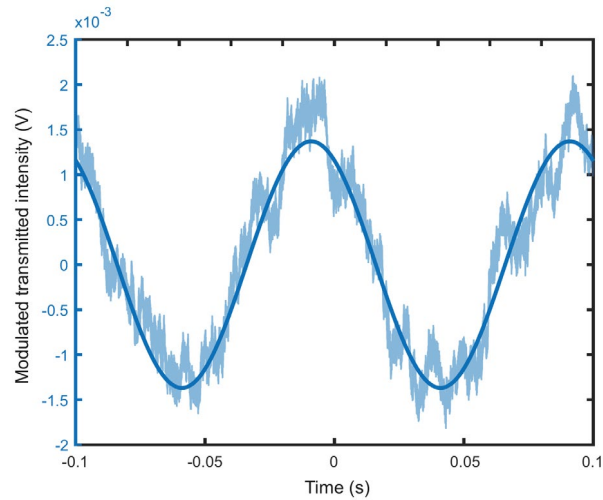

*Figure S12: Electro-optic modulation of the transmission at 10 Hz driving frequency* Measured modulation of the transmitted intensity of the metasurface (radius 135 nm) in time at 775.2 nm for 10 Hz ac voltage frequency, measured with the oscilloscope. Light blue curve represents raw data with bold blue line as a fit.

For the working range of the ac voltage frequency, we detect a measurable change between 10 Hz and 10 MHz. Although even higher frequencies were observed, the electric setup used is not calibrated beyond 10 MHz and therefore no safe statement is possible above that frequency. A shift of the resonance by application of a dc field was not observed for voltages between -10 and 10 V. Using higher voltages would make this shift detectable.

## S12 Fabrication process

The sample layer stack consisted of a lithium niobate substrate (LN, $LiNbO_3$, 500 µm) from NanoLN, a silicon dioxide buffer layer ($SiO_2$, 2 µm), and a LN thin-film (500 nm). We deposited a bi-layer mask of $SiO_x$ (700nm) and chromium (Cr, 100nm) using plasma-enhanced chemical vapour deposition (PECVD) and physical vapour deposition (PVD), respectively.

For electron-beam lithography and development, we used the AR-N 7520.073 resist (217 nm) and the AR300-47 developer. We then dry-etched the Cr and $SiO_x$ mask layers with standard $Cl_2$- and $C_4F_8$-based recipes in the inductively coupled plasma reactive ion etching (ICP-RIE) equipment with the etching rates of 46 nm/min and 282 nm/min, respectively. In the same ICP-RIE, we performed the LN dry-etching with $Ar^+$ and an etching rate of 55 nm/min.

The remaining Cr mask was removed with a commercially available chromium etchant for 2 min, while the SiO$_x$ remaining mask was removed by buffered hydrofluoric acid (BHF) for 4 min (both at room temperature). The LN redeposition was removed by potassium hydroxide (KOH) for 30 min at 70 °C. After the mask removal, we annealed the sample at 250 °C for 1 h to fully recover the LN crystal structure.

We fabricated the electrode structures with direct laser writing on a bilayer resist (1 μm of LOR10B and 500 nm of AZ1505). After development, 10 nm of titanium were deposited followed by 300 nm of gold by electron-beam evaporation. After the lift-off process, the electrodes were connected to the PCB sample holder with silver paint.